\documentclass[showpacs,showkeys,amsmath,amssymb,preprint,nofootinbib]{revtex4-1}
\usepackage{graphicx}
\usepackage{epsfig}
\usepackage{bm}
\usepackage{float}
\usepackage[T1]{fontenc}
\usepackage[utf8]{inputenc}
\usepackage{graphicx}
\usepackage{bm}
\usepackage{amsmath}
\usepackage{xcolor}
\usepackage{comment}
\DeclareUnicodeCharacter{2212}{\textendash}
\def\beq{\begin{eqnarray}}

\def\eeq{\end{eqnarray}}

\usepackage[pdftex, pdftitle={Article}, pdfauthor={Author}]{hyperref}

\usepackage{scalerel}
\usepackage{tikz}
\usetikzlibrary{svg.path}

\definecolor{orcidlogocol}{HTML}{A6CE39}
\tikzset{
  orcidlogo/.pic={
    \fill[orcidlogocol] svg{M256,128c0,70.7-57.3,128-128,128C57.3,256,0,198.7,0,128C0,57.3,57.3,0,128,0C198.7,0,256,57.3,256,128z};
    \fill[white] svg{M86.3,186.2H70.9V79.1h15.4v48.4V186.2z}
                 svg{M108.9,79.1h41.6c39.6,0,57,28.3,57,53.6c0,27.5-21.5,53.6-56.8,53.6h-41.8V79.1z M124.3,172.4h24.5c34.9,0,42.9-26.5,42.9-39.7c0-21.5-13.7-39.7-43.7-39.7h-23.7V172.4z}
                 svg{M88.7,56.8c0,5.5-4.5,10.1-10.1,10.1c-5.6,0-10.1-4.6-10.1-10.1c0-5.6,4.5-10.1,10.1-10.1C84.2,46.7,88.7,51.3,88.7,56.8z};
  }
}

\newcommand\orcidicon[1]{\href{https://orcid.org/#1}{\mbox{\scalerel*{
\begin{tikzpicture}[yscale=-1,transform shape]
\pic{orcidlogo};
\end{tikzpicture}
}{|}}}}

\begin{document}

\title[DEA]{Correspondence between a decaying dark matter sector scenario and scalar field model}

\author{Gilberto Aguilar-P\'erez\orcidicon{0000-0001-6821-4564}$^{1,}$\footnote{gilaguilar@uv.mx}, Deryan Alvarado\orcidicon{0009-0006-2115-0317}$^{2,}$\footnote{zS23000467@estudiantes.uv.mx} Miguel Cruz\orcidicon{0000-0003-3826-1321}$^{1,}$\footnote{miguelcruz02@uv.mx}, Estefany Ru\'\i z-Ramos\orcidicon{0009-0008-4968-9108}$^{1,}$ \footnote{zS19024079@estudiantes.uv.mx}, and Joel Saavedra\orcidicon{0000-0002-1430-3008}$^{3,}$\footnote{joel.saavedra@pucv.cl}}

\affiliation{$^{1}$Facultad de F\'{\i}sica, Universidad Veracruzana 91097, Xalapa, Veracruz, M\'exico \\
$^{2}$Facultad de Matem\'aticas, Universidad Veracruzana 91097, Xalapa, Veracruz, M\'exico \\
$^{3}$Instituto de F\'\i sica, Pontificia Universidad Cat\'olica de Valpara\'\i so, Casilla 4950, Valpara\'\i so, Chile}

\date{\today}

\begin{abstract}
We explore the theoretical viability of modeling a decaying dark matter sector through a unified scalar field approach. Using exact analytical solutions of the Friedmann constraints, we map the fluid phenomenology onto a scalar field potential. Our analysis reveals that physical viability, specifically the existence of a well-defined potential minimum; inevitably forces the dark energy equation of state into the phantom domain. To resolve the kinetic pathologies at late times, we propose reinterpreting the framework within a complex scenario, mapping the imaginary transition to the angular dynamics of a $U(1)$ phase. This mapping naturally yields an ultra-light mass scale of $m_\phi \sim 10^{-33} \ \text{eV}$, classifying the model as a unified dark fluid. Finally, we employ a dynamical approach to study the effects of non-minimal coupling, proving that the phantom-dominated epoch acts as a stable, late-time cosmic attractor in this kind of cosmological scenario.
\end{abstract}



\maketitle

\section{Introduction}
\label{sec:intro}
Understanding the nature of the dark sector remains one of the central challenges in modern cosmology. Observations across multiple scales -- from galactic rotation curves to the cosmic microwave background (CMB) anisotropies -- consistently indicate that approximately 95\% of the energy content of the Universe is composed of non-baryonic components, commonly referred to as dark matter (DM) and dark energy (DE). Despite the remarkable success of the $\Lambda$CDM model in describing the large-scale structure of the Universe, several tensions and small-scale discrepancies have motivated the exploration of alternative scenarios.\\
Among these, scalar field dark matter (SFDM) models have emerged as a compelling framework. In these models, dark matter is described by a fundamental scalar field whose macroscopic behavior can reproduce the phenomenology of cold dark matter at cosmological scales while potentially addressing small-scale issues such as the cusp-core problem and the suppression of substructure formation \cite{Matos2001, Suarez2014, Marsh2016}. In particular, ultra-light scalar fields can exhibit wave-like effects on galactic scales, leading to characteristic features in the matter power spectrum and halo density profiles.\\
On the other hand, increasing attention has been devoted to scenarios in which the dark sector is not composed of isolated components, but instead involves interactions between different dark constituents. A particularly interesting possibility is the existence of dark radiation (DR), a relativistic component that interacts weakly or exclusively within the dark sector. Such a component is motivated both theoretically and observationally, for instance through its potential contribution to the effective number of relativistic species, $N_{\mathrm{eff}}$ \cite{Ackerman2009, Archidiacono2013}. Models involving interactions between dark matter and dark radiation can lead to observable imprints in the CMB and large-scale structure, modifying the growth of perturbations and the matter power spectrum \cite{Tang2016}.\\
More recently, models incorporating dynamical energy transfer within the dark sector -- including the decay of dark matter into dark radiation -- have been explored as a possible mechanism to alleviate cosmological tensions \cite{McCarthy2023}. In this context, both phenomenological and field-theoretical approaches have been developed, ranging from interacting fluid descriptions to scalar-field realizations where the decay process is encoded through effective couplings.\\
Motivated by these developments, in this work we explore a correspondence between a scalar field description and an effective interacting dark sector scenario in which dark matter decays into dark radiation. Our approach is based on constructing an explicit mapping between the scalar field dynamics and an effective fluid description with energy transfer, allowing us to interpret the decay process in terms of the evolution of the scalar degree of freedom and its associated potential. In particular, we analyze the system at the background level by solving the dynamical equations governing the scalar field evolution, focusing on the role of the coupling parameter that controls the energy transfer.\\
We study the behavior of the scalar field as a function of a dimensionless time variable, showing how the interaction modifies its decay rate and dynamical evolution. From this, we reconstruct the effective scalar potential both as a function of time and directly in field space, $V(\phi)$, which provides insight into the underlying structure of the model. The phase-space representation allows us to identify characteristic features of the dynamics, including the impact of the interaction strength on the steepness and curvature of the potential. Additionally, we explore how different parameter choices affect the global behavior of the system, revealing regimes where the dynamics resemble standard scalar field dark matter and others where the interaction induces significant deviations.\\
In a broader context, it is worth noting that interacting dark sector scenarios can also be interpreted within frameworks where the scalar field is non-minimally coupled, either to gravity or to other components of the cosmic fluid. In such cases, the effective energy exchange can arise from geometric couplings or field-dependent interactions, leading to modified evolution equations that can mimic decay processes at the phenomenological level \cite{Faraoni2004, Amendola2000}. Although in this work we adopt an effective description in terms of an interaction function, our results can be viewed as capturing some of the essential features that would emerge in more fundamental non-minimally coupled theories. This connection suggests that the correspondence developed here may provide a useful bridge between phenomenological interacting models and underlying scalar-tensor or extended gravity frameworks.\\
This paper is organized as follows. In Sec.~\ref{sec:ii}, we introduce the theoretical framework and derive the correspondence between the scalar field and the interacting dark sector description. In Sec.~\ref{sec:dynamical}, we explore the inclusion of a dynamical dark energy sector characterized by a constant parameter state. In Sec.~\ref{sec:obs}, some background and perturbation-level observables are analyzed. Finally, in Sec.~\ref{sec:vii}, we summarize our results and outline possible future directions.  $8\pi G = c = 1$ units will be used in our analysis on the framework of a flat Friedmann–Lemaître–Robertson–Walker (FLRW) spacetime with metric given by $ds^{2} = -dt^{2}+a^{2}(t)\left[dr^{2}+r^{2}(d\theta^{2}+\sin^{2}\theta d\varphi^{2}) \right]$ with $a(t)$ being the scale factor, as usual.
 
\section{Decaying dark matter}
\label{sec:ii}
As a starting point, we consider only two components contributing to the total energy of the universe: a dark matter sector that decays into dark radiation. Within the framework of the FLRW metric, this decaying dark matter scenario is characterized by the following set of equations \cite{Ole2012, Poulin2016, Pandey2020}
\begin{align}
& \dot{\rho}_{\mathrm{m}} + 3H\rho_{\mathrm{m}} = - Q, \label{eq:ddm1}\\
& \dot{\rho}_{\mathrm{dr}} + 3H(1+\omega_{\mathrm{dr}})\rho_{\mathrm{dr}} = Q, \label{eq:ddm2}
\end{align}
where $\rho_{\mathrm{m}}$ denotes the dark matter energy density and $\rho_{\mathrm{dr}}$ corresponds to the dark radiation generated by dark matter decay. As usual, we have considered a barotropic EoS between the pressure and density of the fluids. Notice that we are restricting ourselves to the cold dark matter case, i.e. $\omega_{\mathrm{m}}=0$. In addition, $Q$ denotes the decay rate for dark matter. In general, the form of the decay rate is given by, 
\begin{equation}
    Q = \Gamma \rho_{\mathrm{m}},
\end{equation}
where $\Gamma = \alpha H$, with $\alpha$ being a positive constant. As mentioned in \cite{V2008}, $\Gamma$ is expected to vary over time but remain constant in space, which accounts for the appearance of $H$ in the decay rate. As can be seen from the previous equations, the total energy density is conserved, while those related to specific species do not. Then, by considering the previous expressions for the decay rate, we can obtain the following analytic solutions for the energy densities of both components in terms of the scale factor as \cite{Ole2012, Pandey2020}
\begin{align}
& \rho_{\mathrm{m}}(a) = \rho_{\mathrm{m,0}}a^{-(3+\alpha)}, \label{eq:ddm3}\\
& \rho_{\mathrm{dr}}(a) = \rho_{\mathrm{dr,0}}a^{-3(1+\omega_{\mathrm{dr}})} + \rho_{\mathrm{m,0}}a^{-3}\frac{\alpha}{\alpha-3\omega_{\mathrm{dr}}}\left[a^{-3\omega_{\mathrm{dr}}}-a^{-\alpha}\right], \label{eq:ddm4}
\end{align}   
where $a\equiv a(t)$. Notice that for $\alpha=0$ the standard behavior for the dark matter sector and radiation is recovered together with $\omega_{\mathrm{dr}}=1/3$. On the other hand, if $\omega_{\mathrm{dr}}=1/3$, the model is parameterized by a single parameter $\alpha$, in this case the dark radiation sector becomes 
\begin{equation}
    \rho_{\mathrm{dr}}(a) = \gamma a^{-4}+\frac{\alpha}{1-\alpha}\rho_{\mathrm{m,0}}a^{-(3+\alpha)},
\end{equation}
where $\gamma$ is a constant defined as $\gamma \equiv \rho_{\mathrm{dr,0}}+(\alpha\rho_{\mathrm{m,0}})/(\alpha-1)$. For a weak coupling between the components, we must have $\alpha \ll 1$; therefore, $\gamma \simeq \rho_{\mathrm{dr,0}}$. If we assume that initially there is no dark radiation, then we must have $\gamma = 0$ and $\alpha/(1-\alpha)\simeq \alpha$. According to these results, the parameter $\alpha$ denotes the fraction of dark matter that decays into dark radiation since $\rho_{\mathrm{dr}}/\rho_{\mathrm{m}} \simeq \alpha$ relates both densities as follows: $\rho_{\mathrm{dr}}=\alpha \rho_{\mathrm{m}}$  \cite{Ole2012, Pandey2020}. Then, for this cosmological model, the Friedmann constraint reads $3H^{2} = (\rho_{\mathrm{m}}+\rho_{\mathrm{dr}})$ which can be approximated as $H\simeq H_{0}\sqrt{\Omega_{\mathrm{m,0}}a^{-(3+\alpha)}}$. We have written our expression in terms of the fractional energy density denoted as $\Omega$ and the Hubble constant $H_{0}$; this implies that the matter sector departs slightly from the standard $a^{-3}$ scaling, possessing a nonzero parameter state given by $\alpha/3$. This type of behavior aligns with generalized dark matter models, which in turn pave the way for the consideration $\omega_{\mathrm{dm}} \neq 0$ \cite{Hu_1998, Kopp_2018, Kopp_2016, ARAYA2026100515, Li2025}. 
From the usual definition for the Hubble parameter $H\equiv \dot{a}/a$, where the dot denotes derivatives w.r.t. cosmic time, we can solve for the scale factor, yielding
\begin{equation}
    a(t)=\left[1+\frac{(3+\alpha)}{2}\sqrt{\Omega_{\mathrm{m,0}}}H_{0}t\right]^{2/(3+\alpha)}, \label{eq:scale1}
\end{equation}
by assigning the initial value of the scale factor $a_{0}$ as unity; setting $\alpha=0$ in the previous result, we obtain the standard proportional relationship $a(t)\propto t^{2/3}$. Then, as a function of time, the Hubble parameter is 
\begin{equation}
    H(t)=H_{0}\sqrt{\Omega_{\mathrm{m,0}}}\left[1+\frac{(3+\alpha)}{2}\sqrt{\Omega_{\mathrm{m,0}}}H_{0}t\right]^{-1}. \label{eq:timehubble}
\end{equation}
We now incorporate a dark energy component characterized by the cosmological constant, resulting in the following Friedmann constraint $H= H_{0}\sqrt{\Omega_{\mathrm{m,0}}a^{-(3+\alpha)}+\Omega_{\Lambda}}$. Solving in the standard way for the scale factor, we obtain
\begin{equation}
    a(t)=\left(\frac{\Omega_{\mathrm{m,0}}}{\Omega_{\Lambda}} \right)^{1/(3+\alpha)}\sinh^{2/(3+\alpha)}\left[\frac{(3+\alpha)}{2}\sqrt{\Omega_{\Lambda}}H_{0}t\right],\label{eq:scalede}
\end{equation}
which recovers the $\Lambda$CDM model for $\alpha=0$. At early times, it behaves as $a(t)\propto t^{2/(3+\alpha)}$, and at late times, the scaling of the de Sitter solution is recovered $a(t)\propto \exp(\sqrt{\Omega_{\Lambda}}H_{0}t)$. As a result, the Hubble parameter can also be represented as a function of time, yielding
\begin{equation}
    H(t) = H_{0}\sqrt{\Omega_{\Lambda}}\coth\left[\frac{(3+\alpha)}{2}\sqrt{\Omega_{\Lambda}}H_{0}t\right],\label{eq:hubblealpha}
\end{equation}
this solution represents a minimal deviation from the $\Lambda$CDM model, given that $\alpha \ll 1$. Such deviations also arise, for instance, in wave dark matter models, where the quantum nature of the scalar field leads to changes in the standard cosmological expansion history \cite{10338}. From equation (\ref{eq:scalede}), it is evident that $a(t=0)=0$ leads to $H(t=0)\rightarrow \infty$, which corresponds to the Big Bang singularity.   

\subsection{First case: Minimal coupling}
\label{sec:iii}
In order to describe the decaying dark matter as a scalar field, $\phi$, we consider the degeneracy condition between models through the Friedmann constraint
\begin{equation}
    3H^{2}(t) = \frac{1}{2}\dot{\phi}^{2}+V(\phi), \label{eq:sf1}
\end{equation}
where we use the solution for the Hubble parameter given in (\ref{eq:timehubble}) and $V(\phi)$ is the scalar potential. In order to preserve the cosmological principle, the scalar field must depend on cosmic time only $\phi\equiv \phi(t)$. Additionally, the dynamics of $\phi$ is governed by the Klein-Gordon equation, which is given as
\begin{equation}
    \ddot{\phi}+3H\dot{\phi}+V'(\phi)=0, \label{eq:sf2}
\end{equation}
where the prime stands for derivatives w.r.t. $\phi$. Solving the systems of equations (\ref{eq:sf1}) and (\ref{eq:sf2}) together with (\ref{eq:timehubble}), we can write the following expression for the scalar field
\begin{equation}
     \phi(t) = \phi_{0}\mp \frac{2}{\sqrt{3+\alpha}}\ln\left(1+\frac{H_{0}\sqrt{\Omega_{\mathrm{m,0}}}}{2}(3+\alpha)t \right), \label{eq:solution1}
\end{equation}
where $\phi_{0}$ is its initial value, the scalar potential turns out to be
\begin{equation}
    V(\phi) = \frac{H^{2}_{0}\Omega_{\mathrm{m,0}}}{2}(3-\alpha)\exp \left[\pm \sqrt{3+\alpha}\left(\phi-\phi_{0} \right) \right].\label{eq:pot1}
\end{equation}
Our results confirm a known attribute of the exponential potential: the evolution of the scalar field mirrors that of the dominant component. The adoption of exponential and quadratic potentials as descriptions of dark matter has been a key focus in scalar field dark matter models, with $\phi^{2}$ potentials, in particular, shown to replicate the large-scale behavior of the concordance cosmological model \cite{Matos2009}. In our scenario, the universe's evolution is guided by the decaying dark matter sector due to the correspondence condition. Therefore, using (\ref{eq:scale1}) and (\ref{eq:solution1}), we derive $\rho_{\phi} \propto a^{-(3+\alpha)}$. Our scalar field solutions are valid for $t\geq 0$ together with $\phi_{0}>0$; the scalar field solution ($-$) can evolve from positive to negative values as time increases, depending on the value $\phi_{0}$. Specifically, the potential ($-$) does not have a minimum, making this scalar field solution inadequate for describing the dark matter scenario from this perspective. An essential requirement is that the scalar field potential possesses a minimum at some critical value $\phi_{c}$, which permits assigning a non-zero mass scale $m_{\phi}$ to the bosonic particle via the relation $m_{\phi}^{2}=V''(\phi_{c})$ \cite{10urena}. It is important to emphasize that, although dark matter decay is assumed to be small, the instability of the scenario manifests in this alternative formulation through the scalar field potential. For the $(+)$ potential, a global minimum at $\phi = 0$ can exist if the domain of the function is constrained to $\phi \geq 0$. Nevertheless, due to the intrinsic properties of the scalar field, we cannot guarantee that this constraint is always fulfilled.

If we now consider the contribution of the cosmological constant and apply the method again using (\ref{eq:hubblealpha}), we obtain the following expression for the scalar field potential
\begin{equation}
    V(\phi)=-\frac{H_0^2 \Omega _{\Lambda } \left\lbrace \alpha -3 \cosh \left[4 \tanh ^{-1}\left(\exp[\frac{1}{2} \sqrt{3+\alpha} (\phi -\phi_{0})]\right)\right]\right\rbrace}{\cosh \left[4 \tanh ^{-1}\left(\exp[\frac{1}{2} \sqrt{3+\alpha} (\phi -\phi_{0})]\right)\right]-1},
\end{equation}
which can be approximated as
\begin{eqnarray}
    V(\phi) &\simeq & 3H_0^2 \Omega _{\Lambda }\left\lbrace \frac{\cosh \left[4 \tanh ^{-1}\left(\exp[\frac{1}{2} \sqrt{3+\alpha} (\phi -\phi_{0})]\right)\right]}{\cosh \left[4 \tanh ^{-1}\left(\exp[\frac{1}{2} \sqrt{3+\alpha} (\phi -\phi_{0})]\right)\right]-1} \right\rbrace, \nonumber \\
    &=& \frac{3}{4}H_0^2 \Omega _{\Lambda }\left( 3+\cosh\left[\sqrt{3+\alpha} (\phi -\phi_{0}) \right]\right),\label{eq:cosh}
\end{eqnarray}
and has a minimum value at $\phi=\phi_{0}$ given by $3H_0^2 \Omega _{\Lambda }$. For the case in which $\left|\phi-\phi_{0}\right|\gg 1$, the potential (\ref{eq:cosh}) behaves as $V(\phi)\simeq (3/4)H^{2}_{0}\Omega_{\Lambda}\left(3+(1/2)\exp[\sqrt{3+\alpha}\left|\phi-\phi_{0}\right|] \right)$ and for $\left|\phi-\phi_{0}\right|\ll 1$ we obtain $V(\phi)\simeq 3H^{2}_{0}\Omega_{\Lambda}\left(1+(1/8)(3+\alpha)(\phi-\phi_{0})^{2}\right) \simeq 3H^{2}_{0}\Omega_{\Lambda}(1+(3/8)(\phi-\phi_{0})^{2})$. This asymptotic behavior renders the $\cosh$ potential a strong candidate for effectively acting as cold dark matter in the late-time universe. As a result, when a dark energy component is added to the decaying dark matter framework, the whole setup can be consistently described in a unified manner by a single scalar field, with the dark matter behavior encoded in the scalar field potential. The appearance of the $\cosh$ potential in this unified framework is significant, as this specific functional form was pioneered in \cite{Matos01, Alcubierre2002} to resolve small-scale problems in galactic structure formation.

\section{Dynamical dark energy}
\label{sec:dynamical}
In this section, we extend our formulation by including a simple form of a dynamical dark energy sector characterized by a constant parameter state $\omega$, which can take the values $\omega < -1/3$ in order to achieve an accelerated universe. Therefore, $\rho_{\mathrm{de}}(a) = \rho_{\mathrm{de},0}a^{-3(1+\omega)}$ \cite{PhysRevD4439} being $\rho_{\mathrm{de},0}$ the value of the dark energy density at present time, $t_{0}$. The Friedmann constraint reads
\begin{equation}
    H = H_0 \sqrt{\Omega_{\mathrm{m},0} a^{-(3+\alpha)} + \Omega_{\mathrm{de},0} a^{-3(1+\omega)}},\label{eq:dynamical}
\end{equation}
from this point forward, we will formulate everything in terms of the scale factor rather than as explicit functions of time. On the other hand, if we assume that the total cosmic fluid can be described by a canonical scalar field $\phi(t)$ with potential $V(\phi)$. The energy density and pressure for a minimally coupled scalar field are given by 
\begin{equation}
\rho_\phi = \frac12\dot\phi^2+V(\phi),\qquad p_\phi = \frac12\dot\phi^2-V(\phi).
\end{equation}
The Friedmann constraint and the continuity equation for the scalar field read 
\begin{equation}
3H^2=\rho_\phi,\qquad \dot\rho_\phi+3H(\rho_\phi+p_\phi)=0,\label{eq:conse}
\end{equation}
leading to
\begin{equation}
\dot\phi^2 = \rho_{\phi}(1+\omega_{\mathrm{eff}}) = 3H^2(1+\omega_{\mathrm{eff}}),\qquad
V = \frac12(\rho_\phi-p_\phi) = \frac32 H^2(1-\omega_{\mathrm{eff}}),\label{eq:20}
\end{equation}
with $\omega_{\mathrm{eff}}=p_\phi/\rho_\phi$ being the effective parameter state. If we consider the degeneracy condition with the cosmological model (\ref{eq:dynamical}) together with the continuity equation (\ref{eq:conse}), the effective parameter state takes the form
\begin{equation}
1+\omega_{\mathrm{eff}} = \frac{(3+\alpha)\Omega_{\mathrm{m},0} a^{-(3+\alpha)}+3(1+\omega)\Omega_{\mathrm{de},0} a^{-3(1+\omega)}}
{3\bigl(\Omega_{\mathrm{m},0} a^{-(3+\alpha)}+\Omega_{\mathrm{de},0} a^{-3(1+\omega)}\bigr)}.
\end{equation}
If we derive the scalar field with respect to the scale factor, from (\ref{eq:20}) we obtain
\begin{equation}
\frac{d\phi}{da}= \frac{\sqrt{3H^2(1+\omega_{\mathrm{eff}})}}{aH}
= \frac{1}{a}\sqrt{\frac{(3+\alpha)\Omega_{\mathrm{m},0} a^{-(3+\alpha)}+3(1+\omega)\Omega_{\mathrm{de},0} a^{-3(1+\omega)}}
{\Omega_{\mathrm{m},0} a^{-(3+\alpha)}+\Omega_{\mathrm{de},0} a^{-3(1+\omega)}}}.\label{eq:der}
\end{equation}
Therefore, using the previous results, the scalar field potential can also be written in terms of the scale factor as
\begin{equation}
V(a)=\frac12 H_0^2\Bigl[(3-\alpha)\Omega_{\mathrm{m},0} a^{-(3+\alpha)}+3(1-\omega)\Omega_{\mathrm{de},0} a^{-3(1+\omega)}\Bigr].\label{eq:pot}
\end{equation}
Considering the substitution $x=a^{n}$ with $n=\alpha-3\omega$ ($\omega \neq -1$), the integral for $\phi(a)$ can be performed straightforwardly, yielding
\begin{equation}
\phi(a)=\phi_0+\frac{1}{n}\sqrt{3(1+\omega)}\;
\operatorname{arctanh}\!\left(\sqrt{\frac{(3+\alpha)\Omega_{\mathrm{m},0}+3(1+\omega)\Omega_{\mathrm{de},0} a^{n}}{3(1+\omega)\bigl(\Omega_{\mathrm{m},0}+\Omega_{\mathrm{de},0} a^{n}\bigr)}}\right), \label{eq:phi}
\end{equation}
for the case $\omega=-1$ one obtains
\begin{equation}
\phi(a)=\phi_0+\frac{1}{\alpha+3}\operatorname{arcsinh}\!\left(\sqrt{\frac{\Omega_{\mathrm{de},0}}{\Omega_{\mathrm{m},0}}}\,a^{(\alpha+3)/2}\right).
\end{equation}
As can be seen, equations (\ref{eq:pot}) and (\ref{eq:phi}) provide a parametric representation $(\phi(a),V(a))$ of the scalar field potential. Extrema of $V(\phi)$ satisfy $dV/d\phi=0$. Since $d\phi/da\neq0$ (except at isolated points), as can be seen from (\ref{eq:der}); this is equivalent to $dV/da=0$. The derivative of (\ref{eq:pot}) with respect to the scale factor is 
\begin{equation}
\frac{dV}{da}= \frac12 H_0^2\Bigl[-(3+\alpha)(3-\alpha)\Omega_{\mathrm{m},0} a^{-(4+\alpha)}-9(1-\omega^2)\Omega_{\mathrm{de},0} a^{-4-3(1+\omega)}\Bigr].
\end{equation}
then the condition for the extremum is found by multiplying the previous expression by $a^{4+3(1+\omega)}>0$ and using the relation $3(1+\omega)-(3+\alpha)=3\omega-\alpha=-n$, which gives
\begin{equation}
(9-\alpha^2)\Omega_{\mathrm{m},0} a^{-n}+9(1-\omega^2)\Omega_{\mathrm{de},0}=0 \quad\Longrightarrow\quad a^{-n}=-\frac{9(1-\omega^2)}{(9-\alpha^2)}\frac{\Omega_{\mathrm{de},0}}{\Omega_{\mathrm{m},0}}.\label{eq:min}
\end{equation}
For a real, physical solution with $a > 0$ to exist, under the assumption of weak coupling $\alpha \ll 1$, the term $(9-\alpha^2)$ remains strictly positive. Consequently, the term $(1-\omega^2)$ must be negative. This requirement strictly constrains the dark energy equation of state to the phantom regime, $\omega < -1$. Thus, the mathematical potential $V(a)$ possesses a global minimum if and only if the dark energy sector exhibits phantom behavior. While the condition $dV/da = 0$ geometrically locates the minimum, the derivative $d\phi/da$ provides crucial information regarding the physical domain of validity. For the model to correctly represent a canonical scalar field, the kinetic energy must be non-negative ($\dot{\phi}^2 \ge 0$), which enforces the condition $(1+\omega_{\text{eff}}) \ge 0$ and, equivalently, $(d\phi/da)^2 \ge 0$. From equation (\ref{eq:der}), the squared derivative with respect to the scale factor is:
\begin{equation}
    \left(\frac{d\phi}{da}\right)^2 = \frac{3(1+\omega_{\text{eff}})}{a^2} = \frac{1}{a^2} \left[ \frac{(3+\alpha)\Omega_{\text{m},0}a^{-(3+\alpha)} + 3(1+\omega)\Omega_{\text{de},0}a^{-3(1+\omega)}}{\Omega_{\text{m},0}a^{-(3+\alpha)} + \Omega_{\text{de},0}a^{-3(1+\omega)}} \right].
\end{equation}
Evaluating this squared derivative exactly at the minimum $a_{min}$ demonstrates that the product is positive, ensuring the field is real and well-behaved at this specific cosmological epoch. However, as the universe expands ($a \to \infty$), the dark energy term $a^{-3(1+\omega)}$ is inevitably dominating. Because the existence of the minimum explicitly requires $\omega < -1$, the coefficient $3(1+\omega)$ in the numerator is strictly negative. 

Eventually, this negative term governs the overall sign of the numerator, causing $(1+\omega_{\text{eff}})$ to become negative. At this stage, $d\phi/da$ becomes purely imaginary, signaling a physical breakdown of the canonical scalar field framework. Therefore, while the potential exhibits a mathematically valid minimum, the dynamic correspondence dictates that the canonical scalar field will inevitably face a singularity at late times, requiring a transition to a phantom field—characterized by negative kinetic energy—to sustain the $\omega < -1$ expansion regime.

While the transition to a purely imaginary scalar field—signaled by the condition $(d\phi/da)^2 < 0$ at late times—traditionally indicates a pathological breakdown of the canonical framework (typically associated with phantom instabilities), recent advances in cosmological models call for a more refined interpretation. This apparent breakdown serves as a strong theoretical motivation to generalize the framework into a Complex Scalar Field Dark Matter model. By embedding the scenario within a complex manifold, the pathologies associated with the phantom crossing can be circumvented while preserving the virtues of the scalar field correspondence. Let us consider a complex scalar field $\Phi = \frac{1}{\sqrt{2}}\phi e^{i\theta}$ governed by a $U(1)$ symmetric potential. In this generalized scenario, the kinetic term decomposes into a radial component $\dot{\phi}^2$ and an angular component $\phi^2\dot{\theta}^2$. The internal rotational dynamics of the phase $\theta$ effectively contribute to the fluid's equation of state. What manifests as an anomalous imaginary transition for a purely real field is instead mapped to the kinetic energy of the angular degree of freedom in the complex plane. Consequently, the complex scalar field can sustain a highly negative effective pressure ($\omega_{\text{eff}} < -1$) without violating the energy conditions in a way that introduces ghost instabilities. Furthermore, the adoption of a complex scalar field dark matter model provides significant cosmological virtues, particularly regarding structure formation. The internal $U(1)$ symmetry guarantees a conserved Noether charge, which naturally stabilizes the dark matter component. On macroscopic scales, if the mass of the bosonic field is ultra-light, it can undergo a Bose-Einstein Condensation (BEC). The resulting coherent macroscopic state introduces a scale-dependent quantum pressure \cite{complex1, complex2, Complex3}. 

While behaving identically to standard Cold Dark Matter (CDM) on cosmological scales—thus preserving the precise predictions of $\Lambda$CDM for the cosmic microwave background and large-scale structure—this quantum pressure actively suppresses gravitational collapse on sub-galactic scales. This inherent feature provides a dynamical mechanism to resolve the enduring small-scale anomalies of the standard model, such as the core-cusp problem and the missing satellite problem, by naturally smoothing out the central density profiles of dark matter halos.

Therefore, the late-time breakdown observed in the real scalar field correspondence should not be interpreted as a flaw of the underlying physical equivalence. Instead, it indicates that a unified dark sector---where the phase evolution mimics decaying dark matter and the radial potential drives the late-time accelerated expansion---is most robustly described by a complex scalar field. It points toward a broader class of scalar dark matter models that dynamically cross the phantom divide, offering a unified mechanism that could potentially alleviate some cosmological tensions.

Having established that the theoretical framework naturally guides us toward this complex generalization, we now return our attention to the behavior of the field at the potential minimum. The existence of a well-defined, real local minimum is crucial for defining the physical mass of the scalar particle, ensuring the model remains phenomenologically viable during the epochs of structure formation. From (\ref{eq:min}) we obtain the value of the scale factor at the minimum 
\begin{equation}
a_{\min}= \left[\frac{(9-\alpha^2)}{9(\omega^2-1)}\frac{\Omega_{\mathrm{de},0}}{\Omega_{\mathrm{m},0}}\right]^{\!1/n},\qquad n=\alpha-3\omega>0, 
\end{equation}
and the corresponding field value is obtained by inserting our previous result into (\ref{eq:phi}), which in turn results as
\begin{equation}
\phi_{\min}= \phi_0+\frac{1}{n}\sqrt{3(1+\omega)}\;
\operatorname{arctanh}\!\left(\sqrt{\frac{3+\alpha}{3(1+\omega)}}\right).
\end{equation}
For $\omega \lessapprox -1$, the argument of $\operatorname{arctanh}$ exceeds $1$; the analytic continuation via $\operatorname{arctanh}(z)=(1/2)\ln[(1+z)/(1-z)]$ yields a real value. The value of the potential at its minimum is simply
\begin{equation}
V_{\min}= \frac12 H_0^2(3-\alpha)\Omega_{\mathrm{m},0} a_{\min}^{-(3+\alpha)}
\left[1-\frac{3+\alpha}{3(\omega+1)}\right]. 
\end{equation}
Because $\omega + 1 < 0$, the quantity inside the square brackets is positive, which implies that $V_{\min} > 0$. The physical mass is defined by $m_\phi^2 = d^2V/d\phi^2|_{\phi_{\min}}$. Using the chain rule and $dV/da=0$ at the minimum,
\begin{equation}
m_\phi^2 = \frac{d^2V/da^2}{(d\phi/da)^2}\bigg|_{a=a_{\min}}.
\end{equation}
Thus, based on our findings, the mass is given by
\begin{equation}
    m_\phi^2 = \frac{1}{6} H_0^2 (3-\alpha)\,\Omega_{\mathrm{m},0}\, a_{\min}^{-(3+\alpha)}\,
\frac{9\omega^2-\alpha^2}{|1+\omega|},
\end{equation}
which is positive and determined by the dark energy parameter state, the fraction of dark matter that has decayed, $\alpha$, and the present-time dark matter density. To properly contextualize the physical nature of this scalar field, it is highly instructive to perform a numerical estimation of its mass at the potential minimum. Let us assume standard cosmological parameters from recent observations: $\Omega_{\text{m},0} \simeq 0.31$, $\Omega_{\text{de},0} \simeq 0.69$, and $H_0 \simeq 1.44 \times 10^{-33}$ eV in natural units \cite{2020}. Consider a physically viable scenario where the coupling is weak ($\alpha \ll 1$) and the dark energy equation of state is slightly into the phantom regime, parameterized as $\omega = -1 - \epsilon$, where $0 < \epsilon \ll 1$. Under these limits, the exponent is $n = \alpha - 3\omega \simeq 3$. The scale factor at the minimum, $a_{min}$, can be approximated by noting that $\omega^2 - 1 = (-1-\epsilon)^2 - 1 \simeq 2\epsilon$:
\begin{equation}
    a_{min} \simeq \left( \frac{9\Omega_{\text{m},0}}{9(2\epsilon)\Omega_{\text{de},0}} \right)^{1/3} = \left( \frac{\Omega_{\text{m},0}}{2\epsilon \Omega_{\text{de},0}} \right)^{1/3}.
\end{equation}
Consequently, the term in the mass equation scales as $a_{min}^{-(3+\alpha)} \simeq a_{min}^{-3} \simeq 2\epsilon (\Omega_{\text{de},0}/\Omega_{\text{m},0})$. Substituting this approximation back into the exact expression for the effective mass yields a remarkable analytical result. Due to the term $|\omega + 1| = \epsilon$, we can write
\begin{equation}
    m_\phi^2 \simeq \frac{1}{6} H_0^2 (3) \Omega_{\text{m},0} \left( 2\epsilon \frac{\Omega_{\text{de},0}}{\Omega_{\text{m},0}} \right) \frac{9}{\epsilon} \simeq 9 H_0^2 \Omega_{\text{de},0}. \label{eq:mass_approx}
\end{equation}
Taking the square root, the mass of the scalar field at the minimum is inherently fixed by the background cosmology
\begin{equation}
    m_\phi \simeq 3 H_0 \sqrt{\Omega_{\text{de},0}} \simeq 2.49 H_0.
\end{equation}
Given that $H_0 \sim 10^{-33}$ eV, the mass of this scalar field is strictly on the order of $\mathcal{O}(10^{-33})$ eV. Furthermore, a mass of $10^{-33}$ eV does not fall within the traditional Ultra Light/Fuzzy Dark Matter range ($\sim 10^{-22}$ eV). Instead, $m \sim H_0$ corresponds precisely to the characteristic mass scale of Quintessence and dynamical dark energy models. Therefore, rather than representing a pure dark matter particle, this formulation proves that the scalar field acts as a true \textit{Unified Dark Fluid} (often referred to in the literature as Quartessence), see for instance \cite{quart}. The field's ultra-light rest mass ($\sim 10^{-33}$ eV) is fundamentally responsible for driving the late-time cosmic acceleration (acting as dark energy), while its dynamic evolution—encoded in the structure of the potential and its effective equation of state—successfully tracks and mimics the decaying dark matter background at earlier epochs, as commented above.

The minimally coupled scenario provides an exact solution, unequivocally linking the unified dark fluid's mass to the present-day Hubble scale and demonstrating the necessity of the phantom regime for a physically viable minimum. Nevertheless, to evaluate the cosmological robustness of this model, it is necessary to consider more general frameworks where the scalar field couples non-minimally to gravity. In the following section, we depart from the exact analytical integration and adopt a dynamical system approach. As we will show, introducing a non-minimal coupling $\xi R \phi^2$ not only preserves the necessity of the phantom regime ($\epsilon_H < 0$) for the emergence of a minimum, but also provides a formal proof that this state behaves as a stable dynamical attractor, dynamically validating the scalar-fluid correspondence previously derived.

\subsection{Analytical reconstruction approach: non-minimal coupling}
\label{sec:vi}
In this section, we consider the following scalar‑tensor action
\begin{equation}\label{eq:action}
S=\int d^4x\sqrt{-g}\left[\frac{1}{2}F(\phi)R-\frac{1}{2}(\nabla\phi)^2-V(\phi)\right],
\qquad 
F(\phi)=1-\xi\phi^2,
\end{equation}
in a spatially flat FLRW background, the equations of motion obtained from the action (\ref{eq:action}) can be written as \cite{Faraoni2004}
\begin{align}
    & 3H^2 (1 - \xi \phi^2) = \frac{1}{2}\dot{\phi}^2 + V(\phi) + 6\xi H \phi \dot{\phi}, \label{eq:friedmann1} \\
    & -(2\dot{H} + 3H^2) (1 - \xi \phi^2) = \frac{1}{2}\dot{\phi}^2 - V(\phi) - 2\xi \left( \dot{\phi}^2 + \phi \ddot{\phi} \right) - 4\xi H \phi \dot{\phi} \label{eq:friedmann2}, \\
    & \ddot{\phi} + 3H\dot{\phi} + V'(\phi) + 6\xi \left( \dot{H} + 2H^2 \right) \phi = 0, \label{eq:klein_gordon}
\end{align}
observe that setting $\xi = 0$ brings us back to the minimally coupled scenario analyzed in the preceding section. Our aim is to reconstruct $V(\phi)$ \emph{on‑shell}, namely along the classical trajectory, without imposing any specific {\it a priori} functional form for it. We introduce the e‑fold number $N=\ln a$ and the field variable
\begin{equation}
u \equiv \phi,\qquad u' \equiv \frac{du}{dN}.
\end{equation}
Derivatives with respect to $N$ are denoted by a prime, while dots indicate
derivatives with respect to cosmic time $t$:
$\dot\phi = Hu'$, $\ddot\phi = H^2(u''-\epsilon_H u')$, etc. From the degeneracy condition we start from the Hubble parameter $H(a)$ (\ref{eq:dynamical}), one computes the deceleration parameter (or first horizon flow function) \cite{schwarz517higher}
\begin{equation}\label{eq:epsilon}
\epsilon_H \equiv -\frac{H'}{H}
= \frac{3+\alpha}{2}\,s(N) + \frac{3}{2}(1+\omega)\bigl(1-s(N)\bigr),
\end{equation}
where
\begin{equation}
s(N)=\frac{\Omega_{m0}e^{-(3+\alpha)N}}{\Omega_{m0}e^{-(3+\alpha)N}+\Omega_{\mathrm{de0}}e^{-3(1+\omega)N}}.
\end{equation}
For late times ($N\to\infty$) the dark‑energy component dominates,
$s(N)\to0$, and
\begin{equation}\label{eq:epsinfty}
\epsilon_H \,\to\, \frac{3}{2}(1+\omega).
\end{equation}
Hence $\epsilon_H>0$ for $\omega>-1$ (quintessence), $\epsilon_H<0$ for $\omega<-1$
(phantom), and $\epsilon_H=0$ for $\omega=-1$ (cosmological constant).  Both
$H(N)$ and $\epsilon_H(N)$ are treated as known functions. Using the definition given in (\ref{eq:action}), the Friedmann constraint (\ref{eq:friedmann1}) can be written as
\begin{equation}
3F H^2 = \frac12\dot\phi^2+V -3H\dot F,
\end{equation}
where $\dot F = -2\xi H u u'$, which leads to an algebraic form for the potential
\begin{equation}\label{eq:Vexpr}
V = H^2\!\left[3(1-\xi u^2) -6\xi u u' -\frac12 u'^2\right].
\end{equation}
It is convenient to define the dimensionless potential
\begin{equation}
\mathcal{V}(u) \equiv \frac{V}{H^2}
= 3(1-\xi u^2) -6\xi u u' - \frac12 u'^2. \label{eq:Vcal}
\end{equation}
Equation \eqref{eq:Vexpr} already specifies $V$ in terms of the (as yet undetermined) trajectory $u(N)$. To find this trajectory, we require an additional relation connecting $u$ and $u'$, which is supplied by the Klein–Gordon equation (\ref{eq:klein_gordon}). From equations (\ref{eq:Vexpr}) and (\ref{eq:Vcal}), one can write
\begin{equation}\label{eq:Vphi_from_Friedmann_sec4}
V_{,\phi}
= \frac{H^2}{u'}\Big[
-2\epsilon_H \mathcal{V}
-6\xi u u' -6\xi u'^2
-6\xi u u'' - u'u''
\Big].
\end{equation}
On the other hand, the Klein--Gordon equation (\ref{eq:klein_gordon}) gives
\begin{equation}\label{eq:Vphi_from_KG_sec4}
V_{,\phi}
= -H^2\Big[
u'' + (3-\epsilon_H)u' + 6\xi u(2-\epsilon_H)
\Big].
\end{equation}
Equating \eqref{eq:Vphi_from_Friedmann_sec4} and \eqref{eq:Vphi_from_KG_sec4},
and multiplying by $u'/H^2$, leads to the following identity
\begin{equation}\label{eq:G_sec4}
G(u,u') \equiv
(4-6\xi+\epsilon_H)u'^2
+ 6\xi(1+\epsilon_H)u u'
- 6\epsilon_H(1-\xi u^2) = 0.
\end{equation}
Equation \eqref{eq:G_sec4} is not an independent evolution equation but an
\emph{on-shell consistency condition} encoding the compatibility between
the gravitational dynamics and the scalar field equation. It defines a curve in the phase space $(u,u')$ on which all physical trajectories must lie. Consequently, the dynamics reduce to a first-order implicit flow
\begin{equation}
u' = f(u;\epsilon_H).
\end{equation}
Solving the quadratic \eqref{eq:G_sec4} for $u'$ yields the explicit first‑order
flow
\begin{equation}\label{eq:uprime}
u'(u)=\frac{-6\xi(1+\epsilon_H)u
        \pm\sqrt{36\xi^2(1+\epsilon_H)^2u^2
            +24\epsilon_H(4-6\xi+\epsilon_H)(1-\xi u^2)}}
        {2(4-6\xi+\epsilon_H)} .
\end{equation}
The two branches correspond to different physical histories. The discriminant
\begin{equation}\label{eq:Delta}
\Delta(u) \equiv 36\xi^2(1+\epsilon_H)^2u^2 + 24\epsilon_H(4-6\xi+\epsilon_H)(1-\xi u^2)
\end{equation}
must be non‑negative for real $u'$, thereby delimiting the allowed range of $u$.The reconstruction method is inherently parametric: the quantities $H=H(N)$ and $u=u(N)$ are specified as functions of the e-fold number $N$. To obtain an explicit potential $V(\phi)$, one has to invert the relation $u(N)$, which is feasible in regions where $u'\neq 0$. Due to our definitions, $H = H\!\bigl(a(N)\bigr)$.  The nontrivial step is to express $N$ in terms of $u$.  From $u' = du/dN$ we have
\begin{equation}
\frac{dN}{du} = \frac{1}{u'(u)}.
\end{equation}
Integration yields
\begin{equation}\label{eq:Nofu}
N(u) = \int^u \frac{d\tilde u}{u'(\tilde u)} + N_0.
\end{equation}
Since $a = e^N$, we obtain the explicit map
\begin{equation}\label{eq:aofu}
a(u) = a_0\,\exp\!\left( \int^u \frac{d\tilde u}{u'(\tilde u)} \right).
\end{equation}
Because $u=\phi$, this is equivalent to $a(\phi)$.  The potential is then
\begin{equation}
V(\phi) = V\!\bigl(N(\phi)\bigr),
\end{equation}
which is single‑valued wherever $u'\neq 0$.  Points where $u'=0$ mark
breakdowns of local invertibility; $V(\phi)$ may become multi‑valued. Once $a(u)$ is known, the Hubble function can be expressed entirely in terms of
the scalar field:
\begin{equation}\label{eq:Hofphi}
H(\phi) = H\!\bigl(a(\phi)\bigr)
= H\!\left(a_0\,\exp\int^{\phi}\frac{du'}{u'(u')}\right).
\end{equation}
Inserting the explicit $u'(u)$ from \eqref{eq:uprime},
\begin{equation}
\frac{1}{u'(u)} = \frac{2(4-6\xi+\epsilon_H)}
                      {-6\xi(1+\epsilon_H)u \pm \sqrt{\Delta(u)}},
\end{equation}
so that the full cosmological evolution is captured in a single quadrature. This shows that the scalar field does not constitute an independent degree of freedom; its dynamics are entirely determined by the chosen expansion history. 

We now focus on the question of whether stable minima exist. To this purpose, we begin by defining an effective potential of the form derived from (\ref{eq:action})
\begin{equation}\label{eq:Udef}
U(\phi) \equiv V(\phi) - \frac{1}{2}F(\phi)R,
\qquad
R = 6H^2(2-\epsilon_H).
\end{equation}
where we have used the Ricci scalar for a FLRW spacetime, $R=6\left(2H^{2}+\dot{H}\right)$. Using \eqref{eq:Vexpr} one obtains, in dimensionless form,
\begin{equation}\label{eq:UU}
\mathcal{U}(u) \equiv \frac{U}{H^2}
= -3(1-\xi u^2)(1-\epsilon_H) -6\xi u u' -\frac12 u'^2.
\end{equation}
An extremum of $U$ is defined by $dU/d\phi=0$, i.e. $d\mathcal{U}/du=0$.
Because $u'$ is an implicit function of $u$ through $G(u,u')=0$, the total derivative is
\begin{equation}\label{eq:dUdu}
\frac{d\mathcal{U}}{du}
= -6\xi u(1-\epsilon_H) -6\xi(u+u')\frac{du'}{du} - u'\frac{du'}{du},
\end{equation}
and the derivative $du'/du$ is obtained by differentiating the function $G(u,u')$:
\begin{equation}
\frac{\partial G}{\partial u} + \frac{\partial G}{\partial u'}\frac{du'}{du}=0.
\end{equation}
The partial derivatives of $G$ given in \eqref{eq:G_sec4} are
\begin{align}
\frac{\partial G}{\partial u} &= 6\xi(1+\epsilon_H)u' + 12\epsilon_H\xi u, \label{eq:dGdu}\\
\frac{\partial G}{\partial u'} &= 2(4-6\xi+\epsilon_H)u' + 6\xi(1+\epsilon_H)u. \label{eq:dGduprime}
\end{align}
Hence
\begin{equation}\label{eq:duprime}
\frac{du'}{du} = -\,\frac{6\xi(1+\epsilon_H)u' + 12\epsilon_H\xi u}
                       {2(4-6\xi+\epsilon_H)u' + 6\xi(1+\epsilon_H)u}.
\end{equation}
Substituting \eqref{eq:duprime} into \eqref{eq:dUdu} yields a closed algebraic
expression for $d\mathcal{U}/du$. After simplification, this can be written as
a rational function:
\begin{equation}\label{eq:dUdu_rational}
\frac{d\mathcal{U}}{du} = \frac{N(u,u')}{D(u,u')},
\qquad
\mbox{where} \quad D(u,u') = 2(4-6\xi+\epsilon_H)u' + 6\xi(1+\epsilon_H)u.
\end{equation}
The explicit expression of the numerator $N(u,u')$ is a polynomial whose detailed form is not required; the key point is that, on the constraint surface, the condition $d\mathcal{U}/du=0$ is equivalent to $N(u,u')=0$ (because $D$ is generically non‑zero except at isolated points).  Thus, extrema correspond to solutions of the coupled system
\begin{equation}\label{eq:coupled}
\begin{cases}
G(u,u') = 0,\\[2mm]
N(u,u') = 0.
\end{cases}
\end{equation}
To solve the system \eqref{eq:coupled}, we rewrite the expression \eqref{eq:G_sec4} as a quadratic polynomial in $u'$, which leads to the condition $G(u,u')=0$ given by
\begin{equation}\label{eq:quadratic}
A u'^2 + B u u' + C = 0,
\end{equation}
with the coefficients $A = 4-6\xi+\epsilon_H$, $B = 6\xi(1+\epsilon_H)$ and $C = -6\epsilon_H(1-\xi u^2)$. Thus, the discriminant is
\begin{equation}\label{eq:DeltaABC}
\Delta = B^2u^{2} - 4A C.
\end{equation}
Real solutions require $\Delta\ge0$. On the constraint surface, the derivative
$du'/du$ given by \eqref{eq:duprime} can be written as
\begin{equation}\label{eq:duprime_ABC}
\frac{du'}{du} = -\frac{B u' + 2C/u}{2A u' + B u},
\end{equation}
which follows from the identities $\partial G/\partial u = B u' + 12\epsilon_H\xi u$
and $2C = -12\epsilon_H(1-\xi u^2)$; a short calculation confirms the equivalence. Notice that if $u'=0$, the expression given above reduces to $C=0$, i.e.,
\begin{equation}
-6\epsilon_H(1-\xi u^2)=0 \quad\Longrightarrow\quad u_*^2 = \frac{1}{\xi}
\;\;(\epsilon_H\neq0).
\end{equation}
From \eqref{eq:dUdu} with $u'=0$ we have
\begin{equation}
\frac{d\mathcal{U}}{du} = -6\xi u(1-\epsilon_H),
\end{equation}
which vanishes only at $u=0$ (for $\epsilon_H\neq1$). This condition is inconsistent with $u_*^2=1/\xi$ unless $\epsilon_H=0$. Therefore, \emph{no static extremum can occur in a dynamical background} ($\epsilon_H\neq0$); any extrema that do occur must satisfy $u'\neq0$.

Inserting the expression for $du'/du$, Eq. \eqref{eq:duprime_ABC}, into the
extremum condition (\ref{eq:dUdu}),  
we obtain
\begin{equation}
-6\xi u(1-\epsilon_H) - \bigl[6\xi u + (6\xi+1)u'\bigr]
\left(-\frac{B u' + 2C/u}{2A u' + B u}\right) = 0.
\end{equation}
Multiplying by the denominator $2A u' + B u$ and using the constraint
$C = -A u'^2 - B u u'$ to eliminate $C$, a straightforward algebraic simplification
yields a polynomial equation $N(u,u')=0$.  The explicit form of $N$ is
\begin{equation}
N(u,u') = 6\xi u\Bigl[ 2A(1-\epsilon_H)u' + B(1-\epsilon_H)u
           + B u' + \frac{2C}{u} \Bigr] + (6\xi+1)u'\left(B u' + 2\frac{C}{u}\right). \label{eq:Nexpr}
\end{equation}
By inserting $C = -6\epsilon_H(1-\xi u^2)$ and the definitions of $A$ and $B$, one can check that $N(u,u')$ reduces to a third-degree polynomial in $u'$ and $u$. The key observation is that the coupled system \eqref{eq:coupled} is therefore composed of two polynomial equations.

\subsubsection{Non‑existence of minima for $\epsilon_H>0$}
Assume $\epsilon_H>0$ (the usual case $w>-1$). Then $C<0$ for all $|u|<1/\sqrt{\xi}$.
From the quadratic \eqref{eq:quadratic}, $u'$ must be real and nonzero (because
$u'=0$ would imply $C=0$, which cannot occur). Consequently, the discriminant $\Delta>0$
throughout this region. Now examine the extremum condition $N=0$.  Because $C<0$, the term $2C/u$
in $N$ is negative (assuming $u>0$ for definiteness; the other branch is
symmetric).  A detailed sign analysis shows that $N(u,u')$ is strictly
negative when $G=0$ and $\epsilon_H>0$.  This can be proved by substituting
$C = -A u'^2 - B u u'$ into $N$ and simplifying to a manifestly negative
expression.  Consequently, the equation $N=0$ has no real solution on the
physical branch of $G=0$. Hence \emph{no extrema exist} for $\epsilon_H>0$.

\subsubsection{Phantom regime $\epsilon_H<0$ and emergence of minima}
For $\epsilon_H<0$, we have $C>0$ for $|u|<1/\sqrt{\xi}$.  Now the constraint allows turning points (in principle, the case $u'=0$ can occur, although we have already established that it does not correspond to an extremum).  More importantly, the discriminant $\Delta$ can vanish at some $u_*$. At such a point, the two branches of $u'$ merge and $du'/du \to \infty$ (unless the numerator also vanishes, which is not the case
in general).T his peculiar behavior is crucial for the presence of a minimum. We now look for a solution that satisfies both $G=0$ and $\Delta=0$ simultaneously, which correspond to the equations (\ref{eq:quadratic}) and (\ref{eq:DeltaABC}). It follows that
\begin{equation}
B^2 u^2 - 4A C = 0,\qquad A u'^2 + B u u' + C = 0.
\end{equation}
Eliminating $C$ gives $A u'^2 + B u u' + B^2 u^2/(4A)=0$, which factors as
$(2A u' + B u)^2=0$, hence
\begin{equation}\label{eq:Delta0_condition}
2A u' + B u = 0 \quad\Longrightarrow\quad u' = -\frac{B}{2A}u.
\end{equation}
At this value, the denominator in \eqref{eq:duprime_ABC} becomes zero, causing
$du'/du$ to diverge formally. This is exactly the criterion for the effective
potential to have a vertical slope. Evaluating $\frac{d\mathcal{U}}{du}$ at a point satisfying \eqref{eq:Delta0_condition}
requires a limiting procedure. Using the explicit expression \eqref{eq:Nexpr} one can check that the numerator $N(u,u')$ also vanishes when $2A u' + B u = 0$ and the constraint holds. Thus, the point is a candidate for an extremum. To determine the stability we compute the second derivative.  Near a point
where $\Delta=0$, the dominant contribution to $d^2\mathcal{U}/du^2$ comes
from differentiating the term proportional to $du'/du$.  Using the chain rule,
\begin{align}
\frac{d^2\mathcal{U}}{du^2} &=
\frac{\partial}{\partial u}\!\left(\frac{d\mathcal{U}}{du}\right)
+ \frac{\partial}{\partial u'}\!\left(\frac{d\mathcal{U}}{du}\right)\frac{du'}{du}
+ \frac{\partial}{\partial u'}\!\left(\frac{d\mathcal{U}}{du}\right)\frac{d^2u'}{du^2}.
\end{align}
A systematic calculation (best done in the limit $\Delta\to0$) shows that
the leading term behaves as
\begin{equation}\label{eq:d2U_near_Delta0}
\frac{d^2\mathcal{U}}{du^2} \sim -\epsilon_H\,\frac{K(u)}{2A u' + B u},
\end{equation}
where $K(u)>0$ depends on $u$ and $\xi$.  Because the denominator
$2A u' + B u$ is proportional to $\sqrt{\Delta}$ and vanishes, the second
derivative diverges; its sign is determined by $-\epsilon_H$.  For
$\epsilon_H<0$, this gives $d^2\mathcal{U}/du^2 > 0$, indicating a minimum. More precisely, one solves for $u'$ from \eqref{eq:Delta0_condition} and
substitutes back into $G=0$ to find the critical value $u_*$.  The condition
$\Delta(u_*)=0$ together with $C(u_*)>0$ and $A>0$ (typical for $\xi\ll1$) guarantees that the effective mass
\begin{equation}
m_{\mathrm{eff}}^2 = H^2\left.\frac{d^2\mathcal{U}}{du^2}\right|_{u_*},
\end{equation}
is positive and of order $\xi H^2$. Thus, the extremum is a genuine stable minimum.
\subsubsection{The limiting case \(\epsilon_H = 0\)}
The case \(\epsilon_H = 0\) corresponds to an asymptotic de Sitter expansion
(\(\omega = -1\)).  Setting \(\epsilon_H = 0\) in \eqref{eq:G_sec4} we obtain
\begin{equation}
G(u,u')\big|_{\epsilon_H=0} = (4-6\xi)u'^2 + 6\xi u u' = u'\big[(4-6\xi)u' + 6\xi u\big] = 0.
\end{equation}
Hence, the phase space splits into two branches:
\begin{equation}
\text{(i)}\quad u' = 0, \qquad
\text{(ii)}\quad u' = -\frac{6\xi}{4-6\xi}\,u, \label{br2}
\end{equation}
provided \(4-6\xi \neq 0\). The dimensionless effective potential reduces to
\(\mathcal{U}(u) = -3(1-\xi u^2) -6\xi u u' - \frac12 u'^2\).
On branch (i) (\(u'=0\)), \(\mathcal{U} = -3(1-\xi u^2)\), giving an extremum at
\(u_*=0\) with \(d^2\mathcal{U}/du^2 = 6\xi > 0\).  On branch (ii)
(\(u' = -k u\) with \(k = 6\xi/(4-6\xi)\)), \(\mathcal{U}\) becomes a quadratic with a minimum at \(u_*=0\) and positive second derivative for\(0<\xi<2/3\). Thus, the de Sitter limit supports a stable minimum at the origin, where \(F(0)=1\), recalling that $F(\phi)=1-\xi\phi^2$, and the potential acts as an effective cosmological constant.

\subsubsection{Dynamical attractor interpretation}
The first‑order flow $u' = f(u)$ derived from the positive square root in
\eqref{eq:uprime} has a fixed point at $u_*$ satisfying $\Delta(u_*)=0$.
Linearizing around $u_*$, we use \eqref{eq:duprime_ABC} and the fact that
$f'(u)$ near $\Delta=0$ is dominated by the term $1/\sqrt{\Delta}$:
\begin{equation}
f'(u) \simeq -\frac{1}{2A}\frac{\Delta'(u)}{2\sqrt{\Delta(u)}}.
\end{equation}
Choosing the branch that approaches $u_*$ from the allowed region, one finds
$f'(u_*) \to -\infty$, making $u_*$ an attractor.  This establishes the
connection between the algebraic structure of $G=0$ and dynamical stability. These results demonstrate the deep link between the sign of $\epsilon_H$ and the stability of the scalar sector, all derived from the algebraic constraint $G=0$ and its discriminant structure.

\section{The observational signatures and growth of structures}
\label{sec:obs}
\begin{figure}[t]
\centering
\includegraphics[width=0.6\textwidth]{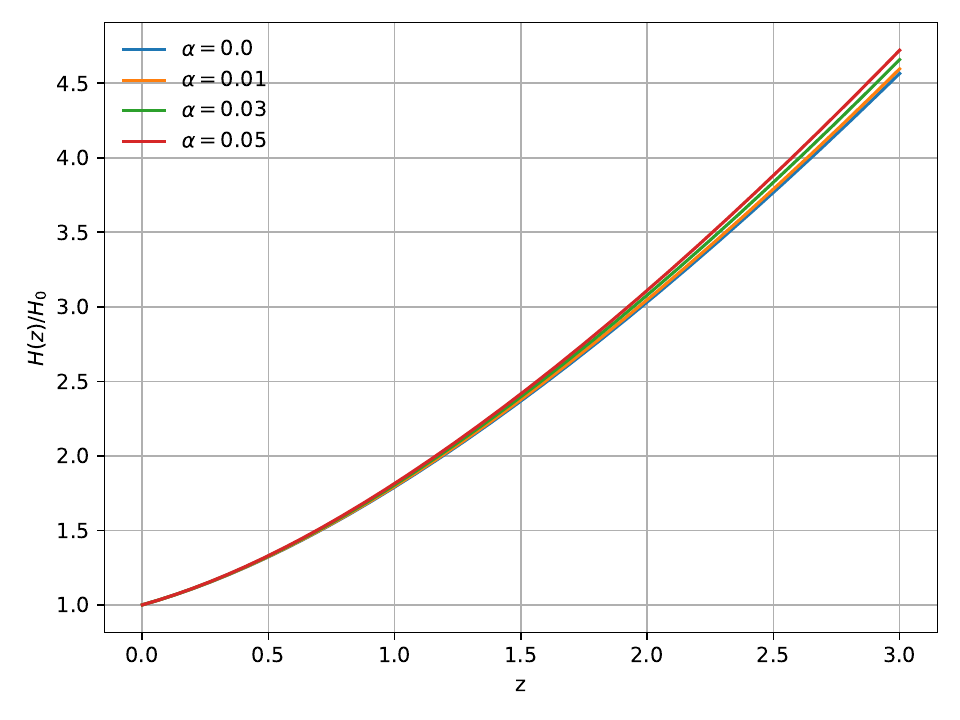}
\caption{
Normalized Hubble expansion rate for different values of the decay
parameter $\alpha$. 
}
\label{fig:Hz}
\end{figure}

\begin{figure}[t]
\centering
\includegraphics[width=0.6\textwidth]{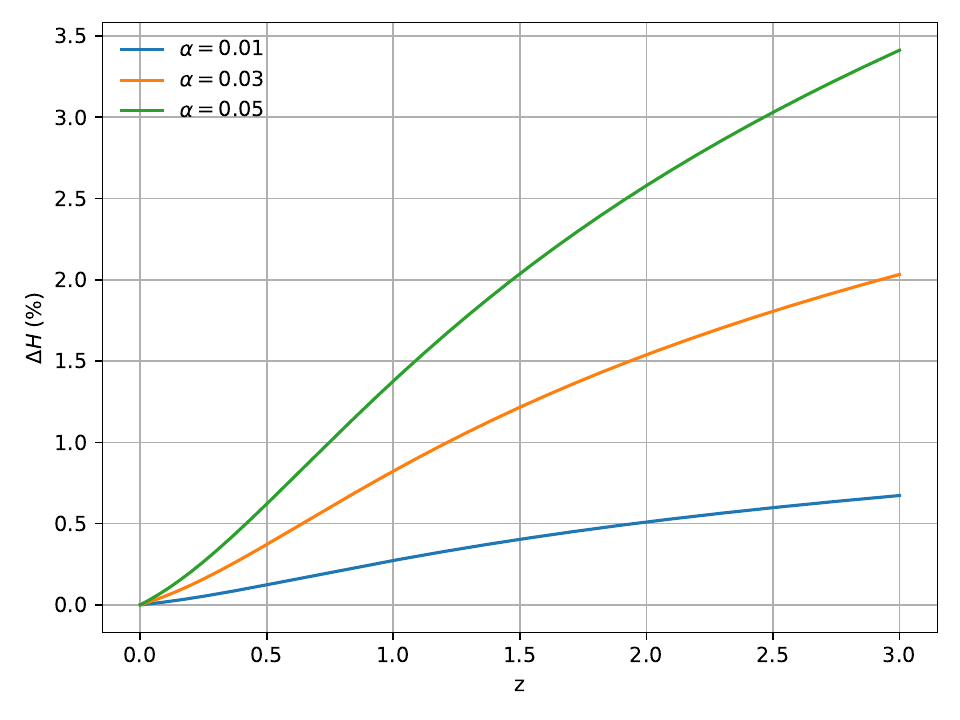}
\caption{
Relative deviation $\Delta_H(z)$ with respect to $\Lambda$CDM.
The effect grows with redshift and with the decay parameter $\alpha$.
}
\label{fig:deltaH}
\end{figure}

\begin{figure}[t]
\centering
\includegraphics[width=0.6\textwidth]{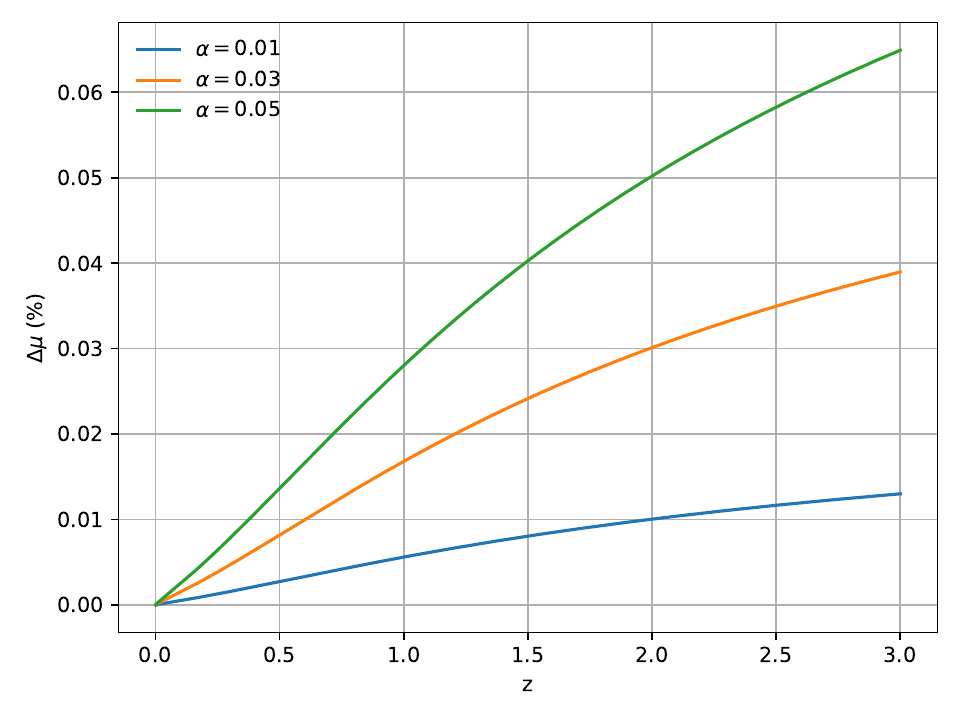}
\caption{
Relative deviation of the distance modulus with respect to $\Lambda$CDM.
}
\label{fig:deltamu}
\end{figure}

\begin{figure}[t]
\centering
\includegraphics[width=0.6\textwidth]{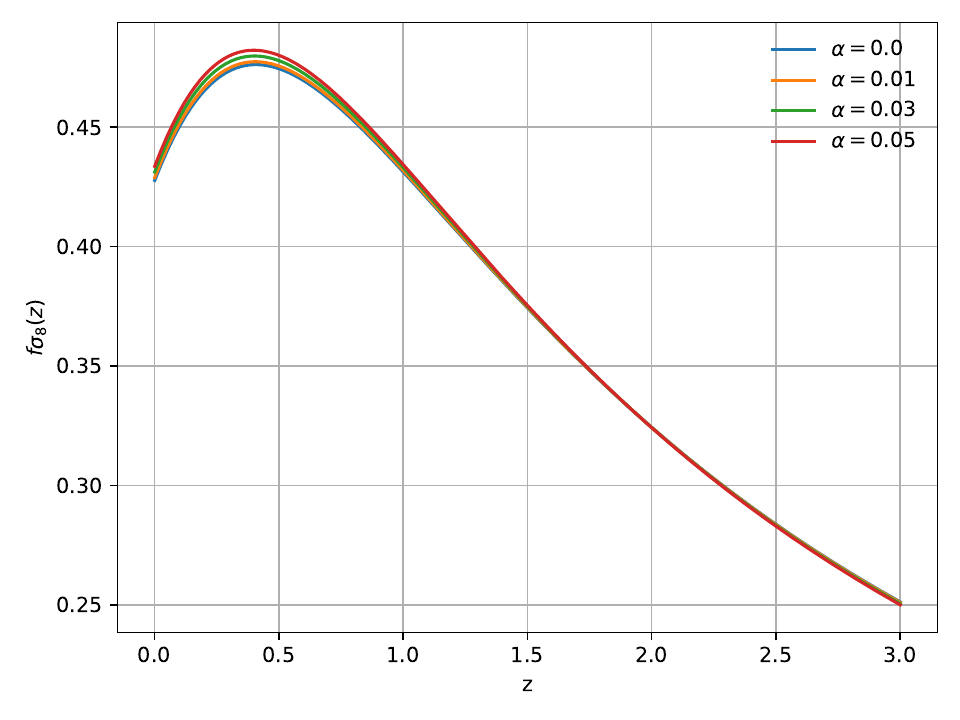}
\caption{
Growth observable $f\sigma_8(z)$ for different values of the decay
parameter $\alpha$. The deviations are small but systematic.
}
\label{fig:fsigma8}
\end{figure}

In order to explore the observational viability of the decaying dark matter
scenario and its scalar-field correspondence, we analyze both the background
and perturbation-level observables. While the degeneracy condition ensures
that the scalar field reproduces the same expansion history, it is crucial
to determine whether this equivalence persists when confronted with
cosmological data.
At the background level, the normalized Hubble function is given by
\begin{equation}
E^2(z) \equiv \frac{H^2(z)}{H_0^2}
=
\Omega_{m,0}(1+z)^{3+\alpha}
+
\Omega_{\Lambda,0},
\end{equation}
where the parameter $\alpha$ quantifies the deviation from the standard
cold dark matter scaling. Figure~\ref{fig:Hz} represents the evolution of $H(z)/H_0$ for several values of $\alpha$. Deviations from the $\Lambda$CDM case
($\alpha=0$) are negligible at low redshift and become progressively
larger at higher redshift.
In order to quantify these deviations, we define the relative difference
\begin{equation}
\Delta_H(z)
=
100\,\frac{|H(z)-H_{\Lambda{\rm CDM}}(z)|}
{H_{\Lambda{\rm CDM}}(z)}.
\end{equation}
 Figure~\ref{fig:deltaH} shows that $\Delta_H$ reaches the percent level
at $z \sim 3$ for $\alpha = 0.05$, indicating that constraints on
$\alpha$ requires high-redshift probes.
The luminosity distance is given by
\begin{equation}
d_L(z)
=
(1+z)\int_0^z \frac{dz'}{H(z')},
\end{equation}
therefore distance modulus reads
\begin{equation}
\mu(z)
=
5\log_{10}\left(\frac{d_L(z)}{\rm Mpc}\right)+25.
\end{equation}
We define the relative deviation
\begin{equation}
\Delta_\mu(z)
=
100\,\frac{|\mu(z)-\mu_{\Lambda{\rm CDM}}(z)|}
{\mu_{\Lambda{\rm CDM}}(z)}.
\end{equation}
Figure~\ref{fig:deltamu} shows that the deviation in $\mu(z)$ remains
extremely small, below $0.1\%$ even for relatively large values of
$\alpha$. This indicates that, at the background level, Type Ia supernova observations are insensitive to the correction due to decaying dark matter.
The degeneracy observed at the background level is broken when considering
the growth of matter perturbations. The evolution of the density contrast
$\delta$ is governed by
\begin{equation}
\ddot{\delta} + 2H\dot{\delta}
- 4\pi G \rho_m\,\delta = 0,
\end{equation}
where the modified matter density scaling
$\rho_m \propto a^{-(3+\alpha)}$ gives us a faster dilution of the matter sector, which reduces the gravitational
clustering efficiency, leading to a suppression of structure formation.
A key observable is the growth rate
\begin{equation}
f(z) = \frac{d\ln \delta}{d\ln a},
\end{equation}
together with
\begin{equation}
f\sigma_8(z) = f(z)\,\sigma_8(z).
\end{equation}
Figure~\ref{fig:fsigma8} shows the evolution of $f\sigma_8(z)$ for
different values of $\alpha$. Although the departures from $\Lambda$CDM are relatively small, they are systematic and offer an independent way to test the model.
Consequently, datasets such as type Ia supernovae are insensitive to $\alpha$, implying that the Hubble expansion yields only weak constraints, while the growth of structure serves as the most powerful probe.
This indicates that the degeneracy between the scalar-field
formulation and the interacting dark sector framework is broken at the
level of perturbations. While both models share the same expansion
history, they predict distinct growth histories, making large-scale
structure observations a powerful discriminator.
Interestingly, the ultra-light mass scale $m_\phi \sim H_0$ implies
that the scalar field behaves as a unified dark fluid, simultaneously
driving cosmic acceleration while modifying the growth of structures.


\section{Concluding remarks}
\label{sec:vii}
In this work, we have established a phenomenological correspondence between a decaying dark matter cosmological scenario and a canonical scalar field framework. By imposing a degeneracy condition on the cosmic expansion history, we analytically reconstructed the scalar field potential and its dynamical evolution, explicitly incorporating a dynamical dark energy component. 

An important outcome of our analysis is the tight bound placed on the dark energy equation of state. We showed that, in order for the scalar field potential to feature a physically viable, real global minimum—required to define the rest mass of the dark matter particle—the dark energy component must be confined to the phantom regime ($\omega < -1$). Furthermore, our analysis of the kinetic domain revealed that the late-time accelerated expansion inevitably drives the squared derivative $(d\phi/da)^2$ to negative values, seemingly violating the canonical energy conditions. However, rather than interpreting this transition as a pathology, we frame it as a strong theoretical motivation to extend the framework into a Complex Scalar Field Dark Matter model. The late-time imaginary transition of the radial field maps naturally to the angular kinetic energy of an internal $U(1)$ phase. This complex generalization not only circumvents instabilities but also provides a dynamic mechanism to stabilize the unified dark sector across cosmic epochs. Finally, the evaluation of the effective scalar mass at the potential minimum yielded a remarkable result. The mass is found to be robustly invariant with respect to the exact magnitude of the phantom parameter, strictly anchoring its scale to the present-day Hubble parameter, $m_\phi \sim \mathcal{O}(10^{-33}) \text{ eV}$. This ultra-light mass scale decisively reclassifies the scalar field. Rather than acting purely as a dark matter particle, the field operates as a true Unified Dark Fluid. Its ultra-light mass drives the late-time cosmic acceleration, while its phase dynamics successfully mimic the decaying cold dark matter background necessary for early structure formation. Crucially, our dynamical system analysis of the non-minimally coupled scenario strengthens these findings. It shows that the appearance of this phantom-dominated minimum is not simply a mathematical byproduct of performing exact analytical integration in the minimally coupled scenario, but instead represents a stable, late-time attractor in the cosmic evolution. This geometric stability ensures that the scalar field can dynamically and naturally settle into the Unified Dark Fluid regime across a broader class of scalar-tensor theories, validating the cosmological viability of the correspondence.

Future observational constraints on the local dark matter density profiles and galactic kinematics will be vital to test the parameter space of this unified complex scalar model, potentially offering a novel, localized method to probe the phantom nature of dark energy.

\section*{Acknowledgments}
M.~Cruz and G. A. P. work has been supported by S.N.I.I. (SECIHTI-M\'exico). G.A.P. was supported by SECIHTI through the {\it Estancias Posdoctorales por México 2023(1)} program. D.~Alvarado  acknowledges SECIHTI doctoral grant. J.~Saavedra acknowledges the FONDECYT grant N°1220065, Chile.

\appendix
\section{Analytical solution for dynamical dark energy}
In this appendix, we address some general aspects of the scenario analyzed in Section \ref{sec:dynamical}. By employing the definition of the Hubble parameter, we obtain
\begin{equation}
    \frac{da}{a \sqrt{\Omega_{\mathrm{m},0} a^{-(3+\alpha)} + \Omega_{\mathrm{de},0} a^{-3(1+\omega)}}} = H_0\,dt,
\end{equation}
whose integral can be written as
\begin{equation}
    \frac{1}{\sqrt{\Omega_{\mathrm{m},0}}} \int a^{\frac{1+\alpha}{2}} \left(1 + \frac{\Omega_{\mathrm{de},0}}{\Omega_{\mathrm{m},0}} a^{\alpha-3w}\right)^{-\frac12} da = H_0 t + C,
\end{equation}
being $C$ an integration constant. If we define $n \equiv \alpha-3\omega$ (assuming $n\neq0$) and perform the change of variable $u = a^{n}$. Then, we can write
\begin{equation}
    \frac{1}{n} \int u^{c-1} \left(1+\frac{\Omega_{\mathrm{de},0}}{\Omega_{\mathrm{m},0}}u\right)^{-\frac12} du,
\end{equation}
where $c = (3+\alpha)/2n$. Compared with the integral \cite{stegun}
\begin{equation}
    \int x^{\alpha}(1-dx^{\beta})^{-\gamma}dx=\frac{x^{\alpha +1}}{\alpha+1}{}_2F_1\left(\gamma, \frac{\alpha+1}{\beta}, \frac{\alpha+1}{\beta}+1; dx^{\beta}\right), \label{eq:app}
\end{equation}
we obtain $\alpha=c-1$, $\beta=1$, $\gamma=1/2$ and $d=\Omega_{\mathrm{de},0}/\Omega_{\mathrm{m},0}$. Imposing $a=0$ at $t=0$ forces $C=0$. Hence, the implicit solution is
\begin{equation}
a^{\frac{3+\alpha}{2}}\; {}_2F_1\!\left(\frac12,\frac{3+\alpha}{2(\alpha-3w)};\frac{3+\alpha}{2(\alpha-3w)}+1;-\frac{\Omega_{\mathrm{de},0}}{\Omega_{\mathrm{m},0}}a^{\alpha-3w}\right)
= \frac{3+\alpha}{2}\sqrt{\Omega_{\mathrm{m},0}}\;H_0 t.
\end{equation}
For $\omega=-1$, $n=\alpha+3$ and $c=1/2$ and using the identity
\begin{equation}
{}_2F_1\!\left(\frac12,\frac12;\frac32;-z\right)=\frac{\operatorname{arcsinh}(\sqrt{z})}{\sqrt{z}},
\end{equation}
the solution reduces to the standard $\Lambda$CDM form with decaying matter discussed before
\begin{equation}
a(t)=\left(\frac{\Omega_{\mathrm{m},0}}{\Omega_{\mathrm{de},0}}\right)^{\frac{1}{3+\alpha}}
\sinh^{\frac{2}{3+\alpha}}\!\left(\frac{3+\alpha}{2}\sqrt{\Omega_{\mathrm{de},0}}\,H_0 t\right).
\end{equation}

\bibliographystyle{ieeetr}
\bibliography{biblio.bib}
\end{document}